\documentclass[11pt]{article}
\usepackage{amsmath}
\usepackage{graphicx,psfrag,epsf}
\usepackage{enumerate}
\usepackage[numbers, sort&compress]{natbib}
\usepackage[hyphens]{url} 

\newcommand{\blind}{0}

\usepackage{makecell}
\usepackage[table,dvipsnames]{xcolor}
\usepackage{amsthm}
\usepackage{bm}
\usepackage{bbm}
\usepackage{mathtools}
\usepackage{collcell}
\usepackage{dsfont}
\usepackage{rotating}
\usepackage{wrapfig}
\usepackage[utf8]{inputenc}
\usepackage{tikz}
\usepackage{lettrine}
\usepackage{dblfloatfix}
\usepackage[english]{babel} 
\usepackage{amsfonts}
\graphicspath{ {fig/} }
\usepackage[hyperfootnotes=false]{hyperref}

\usepackage{subfigure}

\usepackage{lipsum}
\usetikzlibrary{arrows,decorations.markings}
\usepackage{standalone}

\definecolor{gray}{rgb}{0.5,0.5,0.5}
\definecolor{red}{rgb}{0.8,0,0}
\definecolor{dred}{rgb}{0.5,0,0}
\definecolor{blue}{rgb}{0,0.1,1}
\definecolor{dblue}{rgb}{0,0.1,0.6}
\definecolor{cyan}{rgb}{0,0.5,.5}
\definecolor{dcyan}{rgb}{0,0.3,.3}

\definecolor{b}{rgb}{0,0,.8}	
\definecolor{g}{rgb}{0,.6,0}	
\definecolor{n}{rgb}{0,0,0}	
\definecolor{h}{rgb}{0.4,0.2,0.2}	
\definecolor{v}{rgb}{0.2,0.6,0}

\begin{document}

	\def\spacingset#1{\renewcommand{\baselinestretch}%
		{#1}\small\normalsize} \spacingset{1}

	
	\if0\blind
	{
		\title{\bf Probabilistic forecasting of German electricity imbalance prices}
		\author{Michał Narajewski\\
			University of Duisburg-Essen\\
		}
		\maketitle
	} \fi
	
	\if1\blind
	{
		
		\bigskip
		
		\begin{center}
			{\LARGE\bf Probabilistic forecasting of German electricity imbalance prices}
		\end{center}
	
	} \fi
	\begin{abstract}
	The exponential growth of renewable energy capacity has brought much uncertainty to electricity prices and to electricity generation. To address this challenge, the energy exchanges have been developing further trading possibilities, especially the intraday and balancing markets. For an energy trader participating in both markets, the forecasting of imbalance prices is of particular interest. Therefore, in this manuscript we conduct a very short-term probabilistic forecasting of imbalance prices, contributing to the scarce literature in this novel subject. The forecasting is performed 30 minutes before the delivery, so that the trader might still choose the trading place. The distribution of the imbalance prices is modelled and forecasted using methods well-known in the electricity price forecasting literature: lasso with bootstrap, gamlss, and probabilistic neural networks. The methods are compared with a naive benchmark in a meaningful rolling window study. The results provide evidence of the efficiency between the intraday and balancing markets as the sophisticated methods do not substantially overperform the intraday continuous price index. On the other hand, they significantly improve the empirical coverage. The analysis was conducted on the German market, however it could be easily applied to any other market of similar structure.
	\end{abstract}
	
	\noindent%
	\vfill
	
	\newpage
	\spacingset{1.45} 
	
	
	\section{Introduction and motivation}
	
	Since the liberalization of electricity markets the market design has undergone a constant development. Currently, it consists of 3 parts: futures, spot and balancing market. The futures market allows the market participants to trade the electricity in a longer horizon. The spot market consists of the day-ahead and intraday parts, and it is the main electricity market. Here, the market players can trade one day to a few minutes prior the physical delivery. The balancing market, however, is of no less importance as it preserves the system stability. The futures and spot markets are often being run by big energy exchanges, as e.g. the European Energy Exchange (EEX) or Nord Pool, whereas the balancing market is still run locally by the Transmission System Operators (TSOs). Thus, the design of the former ones is rather unified, while the design of the latter one could deviate depending on the control zone. We can particularly distinguish the single and two price imbalance settlement methods. In the study, we consider the German market data, and therefore we focus ourselves on the single price design.
	
	Large deviations from nominal electric grid frequency may lead to disconnections or even blackouts. Thus, the need for electricity balancing is undebatable, and it only gains on importance with the growth of renewable energy capacity, even though the introduction of intraday continuous trading and quarter-hourly products has reduced the need for short-term balancing reserves \cite{ocker2017german, koch2019short}. The German balancing market comprises the capacity and energy markets \cite{Viehmann2017}. The capacity market takes place on the day before the physical delivery period and the traders declare there their balancing capacity for a given price. Then, the balancing service providers (BSPs) that offer the cheapest capacity are accepted and may participate in the balancing energy market. A detailed description of the market is presented in Section~\ref{sec:market}.
	
	This paper raises the novel issue of very short-term probabilistic forecasting of German electricity imbalance prices. We apply the methods well-known in electricity price forecasting (EPF) in order to model and predict the distribution of imbalance prices 30 minutes before the delivery. The motivation for such setting is the possibility to trade the energy in the intraday continuous market in the respective control zones, after gate closure, until 5 minutes before the delivery or in the balancing market. Having precise imbalance price probabilistic forecasts and access to intraday continuous limit order book, the market participant may choose between these two to maximize their profit.
	The utilized modelling methods are: lasso with bootstrapped in-sample errors, gamlss with lasso-based variable selection and probabilistic neural networks. 
	For gamlss and neural networks we assume two distributions: normal and Student's t.
	The models are compared against a naive benchmark -- EPEX ID$_1$ Price in a rolling window study, what is inline with the existing EPF literature. The models are presented in detail in Section~\ref{sec:models} and the application study in Section~\ref{sec:study}.
	
	The electricity balancing markets have already drawn the researchers' attention. The balancing market design was studied by \citet{van2012agent, van2016electricity, poplavskaya2020effect}. The authors additionally analyse the impact of the imbalance pricing mechanism on market behaviour, and they conclude that although the system imbalance is similar for different mechanisms, the mechanism that minimizes the imbalance costs for the market is the single price settlement. The literature on modelling and forecasting in electricity balancing markets can be split to imbalance forecasting \cite{toubeau2021interpretable, bunn2018trading, bottieau2019very, bunn2021statistical}, imbalance price forecasting \cite{klaeboe2015benchmarking, lucas2020price, dumas2019probabilistic} and the application in trading \cite{browell2018risk, bunn2021statistical, bunn2018trading, bottieau2019very, kumbartzky2017optimal}. The scarce electricity imbalance price forecasting literature focuses on point forecasting \cite{klaeboe2015benchmarking, lucas2020price}, interval forecasting \cite{klaeboe2015benchmarking} and probabilistic forecasting \cite{dumas2019probabilistic}. The work of \citet{dumas2019probabilistic} is naturally the closest one to our study. The authors utilize a two-step approach, namely they first calculate the probabilities for the net imbalance and then based on that make predictions regarding the imbalance prices. On the other hand, we forecast the imbalance prices directly and do not make any prior assumptions.
	
	The research on EPF is much wider than the one particularly focused on balancing markets. \citet{weron2014electricity} provides a review of point forecasting methods and \citet{nowotarski2018recent} present an overview of probabilistic forecasting methods in electricity markets. 
	The big majority of the EPF literature considers the day-ahead market 
	\cite{
		ziel2016forecasting, ziel2018day, marcjasz2018selection, lago2018forecasting, serafin2019averaging,  
		lago2021forecasting}, however the intraday market gains on importance both in practice and in literature \cite{uniejewski2019understanding, narajewski2020econometric, oksuz2019neural,janke2019forecasting, narajewski2020ensemble, marcjasz2020beating}.
	Similarly, much more research has been done on point forecasting than on probabilistic forecasting \cite{nowotarski2018recent}. The most popular and effective methods in recent point EPF literature are lasso \cite{ziel2016forecasting, ziel2018day, lago2021forecasting, uniejewski2019understanding, narajewski2020econometric, marcjasz2020beating} and deep neural networks 
	\cite{lago2021forecasting, oksuz2019neural}, whereas for probabilistic EPF we can name quantile regression \cite{maciejowska2016probabilistic, maciejowska2020assessing, uniejewski2021regularized} and gamlss \cite{gianfreda2018stochastic, narajewski2020ensemble}.
	A relatively big attention is also being paid to the forecasting combination of electricity prices
	\cite{nowotarski2015computing, marcjasz2018selection, serafin2019averaging}. 
	
	Now, let us summarize the major contributions of the manuscript:
	\begin{enumerate}
		\itemsep0em
		\item It is the first work on direct probabilistic forecasting of electricity imbalance prices.
		\item The imbalance market is inevitable for any market player, and thus this paper may contribute also to electricity trading literature.
		\item Various probabilistic models are compared in an exhaustive forecasting study.
		\item We contribute to the scarce electricity balancing literature by drawing researchers' attention to the German electricity balancing market.
		\item The paper provides evidence of the efficiency between the intraday and balancing markets.
	\end{enumerate}
	Let us additionally note that the importance of this research is emphasized by the need of including the imbalance market in the electricity trading strategies \cite{narajewski2021optimal}.
	
	The remainder of this manuscript has the following structure. Section~\ref{sec:market} describes the electricity balancing market in Germany, the calculation of the imbalance price and the data utilized in the study. The models and estimation methods are discussed in Section~\ref{sec:models}. Section~\ref{sec:study} presents the application study, including the description of the setting, and the empirical results. Finally, Section~\ref{sec:conclusion} closes the paper with conclusion.
	\section{Electricity balancing market}\label{sec:market}
	This section familiarizes the reader with the German balancing market and provides a description of calculation of the imbalance price. Additionally, we present the data used in the purpose of this study.
	\subsection{Balancing market in Germany}\label{sec:balancing}
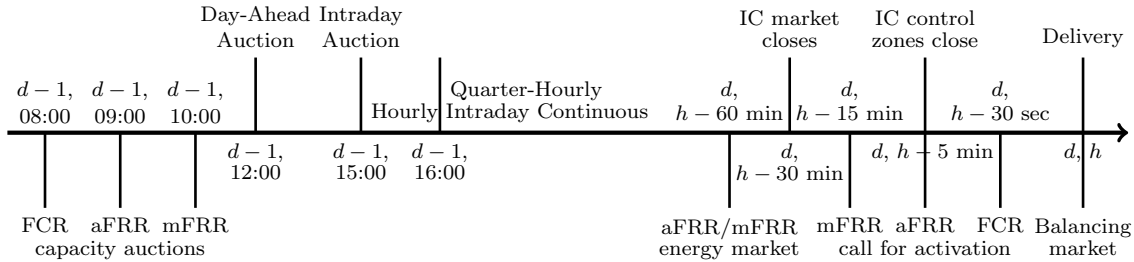
\begin{figure*}[!t]
	\begin{tikzpicture}[scale=1]
		\draw [->] [ultra thick] (0,0) -- (14.9,0);
		\draw [line width = 1] (0.5,0) -- (0.5, -1);
		\node [align = center, above, font = \scriptsize] at (0.5, 0.3) {$d-1$,};		
		\node [align = center, above, font = \scriptsize] at (0.5, 0) {08:00};		
		\node [align = center, below, font = \scriptsize] at (0.5, -1) {FCR};
		\draw [line width = 1] (1.5,0) -- (1.5, -1);
		\node [align = center, above, font = \scriptsize] at (1.5, 0.3) {$d-1$,};	
		\node [align = center, above, font = \scriptsize] at (1.5, 0) {09:00};		
		\node [align = center, below, font = \scriptsize] at (1.5, -1) {aFRR};
		\draw [line width = 1] (2.5,0) -- (2.5, -1);
		\node [align = center, above, font = \scriptsize] at (2.5, 0.3) {$d-1$,};		
		\node [align = center, above, font = \scriptsize] at (2.5, 0) {10:00};
		\node [align = center, below, font = \scriptsize] at (2.5, -1) {mFRR};
		\node [align = center, below, font = \scriptsize] at (1.5, -1.3) {capacity auctions};
		\draw [line width = 1] (3.3,1) -- (3.3, 0);
		\node [align = center, below, font = \scriptsize] at (3.3, 0) {$d-1$,};
		\node [align = center, below, font = \scriptsize] at (3.3, -0.3) {12:00};
		\node [align = center, above, font = \scriptsize] at (3.3,1.3) {Day-Ahead};
		\node [align = center, above, font = \scriptsize] at (3.3,1) {Auction};
		\draw [line width = 1] (4.7,1) -- (4.7, 0);
		\node [align = center, below, font = \scriptsize] at (4.7, 0) {$d-1$,};
		\node [align = center, below, font = \scriptsize] at (4.7, -0.3) {15:00};
		\node [align = center, above, font = \scriptsize] at (4.7,1.3) {Intraday};
		\node [align = center, above, font = \scriptsize] at (4.7,1) {Auction};
		\node [align = center, above right, font = \scriptsize] at (4.7,0) {Hourly Intraday Continuous};
		\draw [line width = 1] (5.75,1) -- (5.75, 0);
		\node [align = center, below, font = \scriptsize] at (5.75, 0) {$d-1$,};
		\node [align = center, below, font = \scriptsize] at (5.75, -0.3) {16:00};
		\node [align = center, above right, font = \scriptsize] at (5.75,0.3) {Quarter-Hourly};
		\draw [line width = 1] (9.6,0) -- (9.6, -1);
		\node [align = center, below, font = \scriptsize] at (9.6,-1) {aFRR/mFRR};
		\node [align = center, below, font = \scriptsize] at (9.6,-1.3) {energy market};
		\node [align = center, above, font = \scriptsize] at (9.6, 0.3) {$d$,};	
		\node [align = center, above, font = \scriptsize] at (9.6, 0) {$h - 60$ min};		
		\draw [line width = 1] (10.4,1) -- (10.4, 0);
		\node [align = center, above, font = \scriptsize] at (10.4,1.3) {IC market};
		\node [align = center, above, font = \scriptsize] at (10.4,1) {closes};
		\node [align = center, below, font = \scriptsize] at (10.4, 0) {$d$,};
		\node [align = center, below, font = \scriptsize] at (10.4, -0.3) {$h - 30$ min};
		\draw [line width = 1] (11.2,0) -- (11.2, -1);
\node [align = center, below, font = \scriptsize] at (11.2,-1) {mFRR};
\node [align = center, above, font = \scriptsize] at (11.2, 0.3) {$d$,};
\node [align = center, above, font = \scriptsize] at (11.2, 0) {$h - 15$ min};
		\draw [line width = 1] (12.2,1) -- (12.2, -1);
		\node [align = center, above, font = \scriptsize] at (12.2,1.3) {IC control};
		\node [align = center, above, font = \scriptsize] at (12.2,1) {zones close};
		\node [align = center, below, font = \scriptsize] at (12.3, 0) {$d$, $h - 5$ min};
	\node [align = center, below, font = \scriptsize] at (12.2,-1) {aFRR};
	\node [align = center, below, font = \scriptsize] at (12.2,-1.3) {call for activation};
		\draw [line width = 1] (13.2,0) -- (13.2, -1);
\node [align = center, below, font = \scriptsize] at (13.2,-1) {FCR};
\node [align = center, above, font = \scriptsize] at (13.2, 0.3) {$d$,};
\node [align = center, above, font = \scriptsize] at (13.2, 0) {$h - 30$ sec};
		\draw [line width = 1] (14.3,1) -- (14.3, -1);
		\node [align = center, above, font = \scriptsize] at (14.3,1) {Delivery};
		\node [align = center, below, font = \scriptsize] at (14.3, 0) {$d$, $h$};
		\node [align = center, below, font = \scriptsize] at (14.3, -1) {Balancing};
		\node [align = center, below, font = \scriptsize] at (14.3, -1.3) {market};
	\end{tikzpicture}
	\caption{The daily routine of the German electricity spot (top) and balancing (bottom) markets. $d, h$~correspond to the day and hour of the delivery, respectively.}
	\label{fig:market}
\end{figure*}

	The balancing market is a crucial part of every electricity market. In Germany, it was adjusted many times in recent years, unlike the spot market, which is already well-developed and the appearing changes are rather minor. The current timeline of electricity spot and balancing market in Germany can be seen in Figure~\ref{fig:market}.
	The spot market is presented in the top part and it consists of the Day-Ahead Auction (DA), Intraday Auction (IA) and Intraday Continuous (IC). The DA takes place on the day before the delivery at 12:00 and it is the main part of the market, where the majority of power volume is traded. The IA takes place 3 hours after the DA, at 15:00 and here the market participants can trade quarter-hourly contracts, whereas in the DA one may trade only hourly contracts. This part of the market serves mainly the purpose of balancing the ramping effects of demand and power generation
	\cite{kremer2020intraday, kremer2021econometric}, however \citet{narajewski2021optimal} show that a trader could make significant gains by incorporating this market in their trading strategy. The IC is the last part of the spot market, and it starts on the day before the delivery at 15:00 for hourly products and at 16:00 for quarter-hourly products\footnote{Strictly speaking, one can trade also half-hourly products starting at 15:30, however they are not very popular in the market.}. Here, the market players can trade power continuously until 30 minutes before the delivery in whole Germany and until 5 minutes before the delivery in respective TSO control zones\footnote{The German market is divided to 4 control zones.}. Also, starting at 22:00 the previous day until 1 hour before the delivery the market participants can trade cross-border using the XBID system \cite{kath2019modeling}. The purpose of the IC market is to enable the traders to react to changing generation or consumption forecasts and adjust their positions. Even though the trading window is very long, the most of the power volume traded in the IC is traded in the last couple of hours before the delivery 
	\cite{narajewski2019estimation, kramer2021exogenous}. Therefore, the most important IC price indicators are the volume-weighted average prices ID$_1$ and ID$_3$ \cite{uniejewski2019understanding, narajewski2020econometric, narajewski2020ensemble} which measure the price level in the last 1 and 3 hours before the delivery, respectively.
	
	For the spot market participants of particular interest is the balancing market and especially the imbalance price. As many of the producers and consumers face high uncertainty due to the stochastic nature of weather conditions and people's behaviour, it is basically impossible for them to balance their generation or consumption perfectly. Thus, any deviations from the scheduled generation or consumption are then handled by the TSOs during the delivery. The costs of balancing the energy are then divided between the market players, often called balance responsible parties (BRPs), who contributed to the imbalance. On the other hand, the BRPs that deviated from their schedule, but their deviation reduced the overall system imbalance, are rewarded for this imbalance reduction. Let us note that even though we name the final energy balancing a market which is inevitable for any market participant, it is not really a market in which the BRPs can make bids. Instead, they need to accept the imbalance price that is a derivative of total balancing costs and total system imbalance.
	
	The bottom part of Figure~\ref{fig:market} presents the balancing market routine. To avoid big deviations from nominal frequency in the electricity grid, the TSOs have three types of BSPs at their disposal: FCR, aFRR and mFRR. The Frequency Containment Reserve (FCR), also referred as primary reserve, is fully activated after 30 seconds and is a first response to any occurring imbalance. If the imbalance persists, the Automatic Frequency Restoration Reserve (aFRR), also referred as secondary reserve, is activated and in case of longer and deeper imbalances, the  Manual Frequency Restoration Reserve (mFRR), also referred as tertiary reserve, is activated. The full activation time of aFRR and mFRR is 5 and 15 minutes, respectively. The balancing market is divided to capacity and energy markets. In the capacity market, the BSPs offer their readiness to deliver or receive the unscheduled electricity and in the energy market, they define the costs for given amount of balancing energy. Let us note that the balancing services are offered in 4-hours positive or negative blocks and the FCR does not participate in the energy market due to negligible volumes.
	
	The capacity auctions take place on the day before the delivery at 08:00 (FCR), 09:00 (aFRR) and 10:00 (mFRR)\footnote{In the past, they were taking place in the week before the delivery, and later also two days before the delivery \cite{Viehmann2017}.}. Based on the demand from TSOs, the cheapest offers are accepted. The winning BSPs are remunerated with pay-as-cleared (FCR) and pay-as-bid (aFRR and mFRR) mechanisms\footnote{In future it is planned to incorporate the pay-as-cleared mechanism also for aFRR and mFRR.}. Then, until one hour before the 4-hour delivery block the BSPs can make bids in the energy market\footnote{In the past, the capacity and energy markets were taking place simultaneously.}. The offers are sorted creating a merit order list and in case of imbalance they are activated with pay-as-bid remuneration mechanism. The costs of balancing energy are carried over to BRPs, whereas the costs of balancing capacity are carried over to end consumers.
	
	\subsection{Imbalance price}
	
	As mentioned, in the German electricity market (but also in many other European markets) the imbalance price is settled using a single price mechanism. The German TSOs have established a Grid Control Cooperation (GCC) and thus the price is unified for all German control zones. The basic formula is as follows
	\begin{equation}
		\text{IP}^{d,qh}_{\text{basic}} = \frac{\sum \text{Costs}^{d,qh}_{GCC} - \sum \text{Revenues}^{d,qh}_{GCC}}{\text{net balance position}^{d,qh}_{GCC}}.
		\label{eq:IP1}
	\end{equation}
	Let us note that the price is in EUR/MWh, and it is calculated separately for each quarter-hour. The balancing costs and revenues of the GCC are derived based on the activated energy from aFRR and mFRR suppliers. Since the numerator and denominator of equation~\eqref{eq:IP1} can be both negative and positive, the same applies to the imbalance price. The BRPs that contribute to the imbalance, i.e. are short/long in case of system under/over-supply, pay the price to the TSOs. However, the BRPs that reduce the system imbalance by being short/long in case of system over/under-supply are being paid the price by the TSOs.
	
	The price given in equation~\eqref{eq:IP1} is not the final imbalance price. Before it reaches its ultimate value, it undergoes multiple modifications. In the following, we list the modifications, however we do not go deep into details as the formulas are cumbersome and not much explanatory.
	\begin{enumerate}
		\itemsep0em
		\item Price cap in the case of a small GCC balance.
		\item Additional price cap in the case of a small GCC balance.
		\item Price comparison with the intraday market and setting a minimum price distance to it in such direction that it is less profitable to contribute to the imbalance.
		\item Surcharge/discount on the imbalance price in the event of GCC reaching 80\% of the positive/negative balancing capacity.
	\end{enumerate}
	The details of the current and past imbalance price calculation method are available on the regelleistung.net website \cite{regelleistung}. The first two modifications are meant to avoid extreme imbalance prices in the case of small net GCC balance. The third one compares the imbalance price with the Intraday Price Index and sets a minimum distance of 25\%, but at least 10 EUR/MWh between them. This modification pushes the price in such direction that the BRPs contributing to the imbalance get worse price as they would have got in the intraday market. The Intraday Price Index is a volume-weighted average price that uses for calculation all the transactions in the intraday continuous on the hourly and quarter-hourly product on the particular day. The fourth modification is an additional penalty on the BRPs that contribute to the system imbalance in the case it reaches very high values. All the measures make it very unprofitable to contribute to the imbalance, but on the other hand very lucrative for the BRPs to reduce it. We denote the adjusted imbalance price as $\text{IP}^{d,qh}$ and refer to it as the imbalance price.
	
\begin{figure}[b!]
	\centering
	\includegraphics[width=\linewidth]{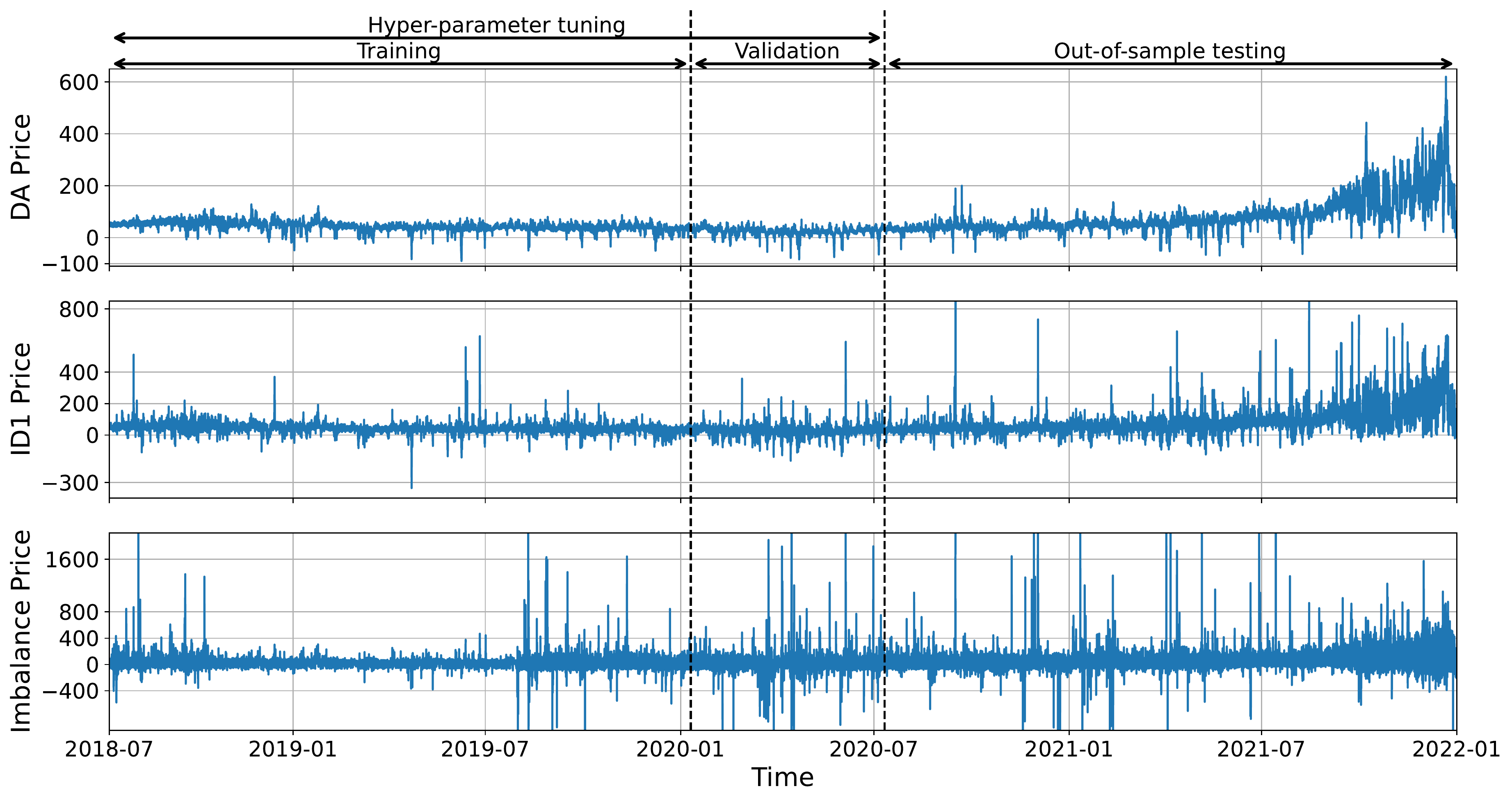}
	\caption{Time series plots of various electricity price data in EUR/MWh.}
	\label{fig:price_tsplot}
	\end{figure}

\begin{figure}[b!]
	\centering
	\includegraphics[width=\linewidth]{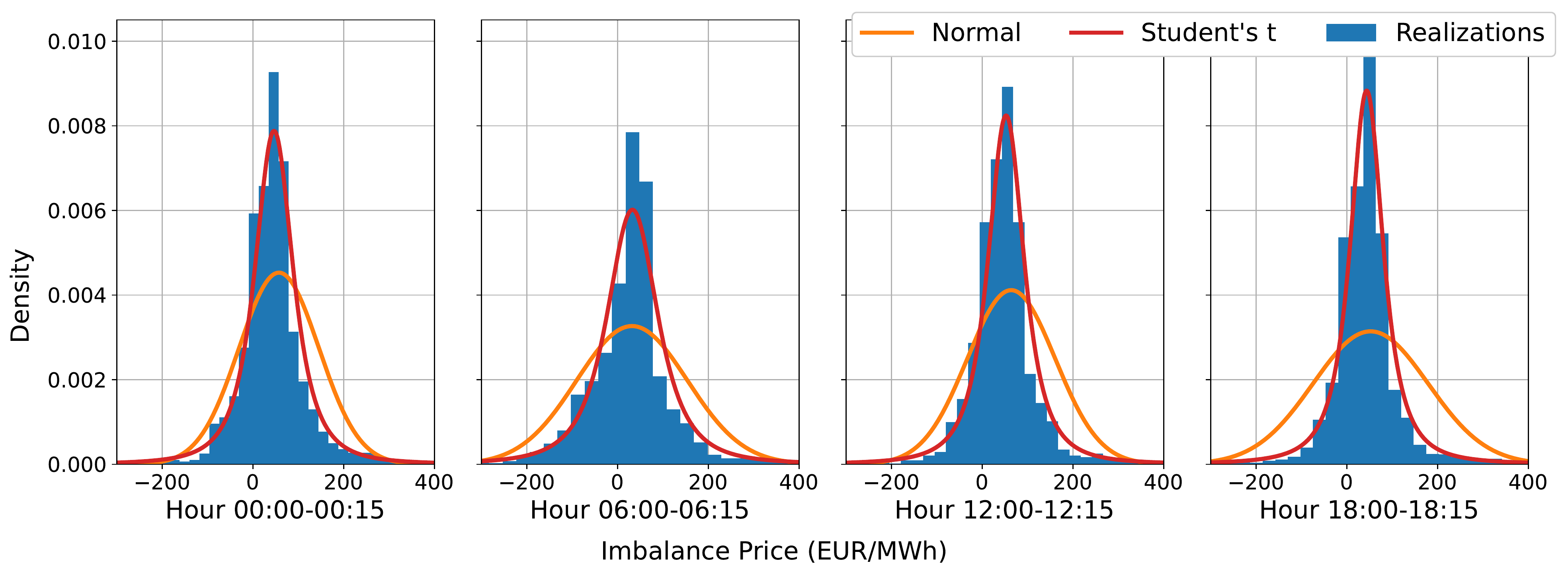}
	\caption{Histograms of imbalance prices with fitted densities for selected hours.}
	\label{fig:price_histogram}
	\end{figure}

	Figure~\ref{fig:price_tsplot} presents the time series of three electricity prices: the DA price, the quarterly ID$_1$ price, and the imbalance price.
	The plots show clearly that the imbalance price is much more volatile than the prices in the quarterly IC market or in the DA market. Moreover, in the considered time-frame the imbalance price exhibited many positive and negative extreme spikes, with a minimum of around $-6500$ EUR/MWh and a maximum of around $24500$ EUR/MWh (for better clarity of Figure~\ref{fig:price_tsplot} we do not show the extremes in the plot).  Such values are impossible to reach in the DA (a min of $-500$ and a max of $3000$) and IC (a min of $-9999$ and a max of $9999$) markets. Therefore, the participation in the balancing market comes with a high risk for a BRP. This confirms Figure~\ref{fig:price_histogram} which shows histograms of imbalance prices for selected hours (the range of prices was limited for better clarity of the histograms).
	The fitted densities prove that the data is heavy-tailed as the three parametric student's t distribution $t(\mu, \sigma, \tau)$ seems to fit the data much better than the normal distribution $\mathcal{N}(\mu, \sigma^2)$. The two mentioned distributions will be later utilized in the application study. All the distributions belong to the location-scale family with $\mu$, $\sigma$, $\tau$ and being the location, scale, and tail-weight (degrees of freedom) parameters, respectively.

	\subsection{Data}
	
	The data utilized in the study are collected from 4 different sources. The spot market data (DA, IA and IC transactions and prices) from the EEX transparency, the day-ahead forecast data (load and renewable generation) from the ENTSO-E transparency, the balancing market data (imbalance price, imbalance volume, aFRR and mFRR capacity and energy market data) from the regelleistung.net and the fuels and emission allowance prices from the ICE. The complete dataset contains observations between 12.07.2018 and 31.12.2021 as the aFRR and mFRR data is not available for the preceding time. We cleaned the data from missing values using the R package tsrobprep \cite{narajewski2021tsrobprep}.
	
\begin{figure}[t!]
	\centering
	\includegraphics[width=\linewidth]{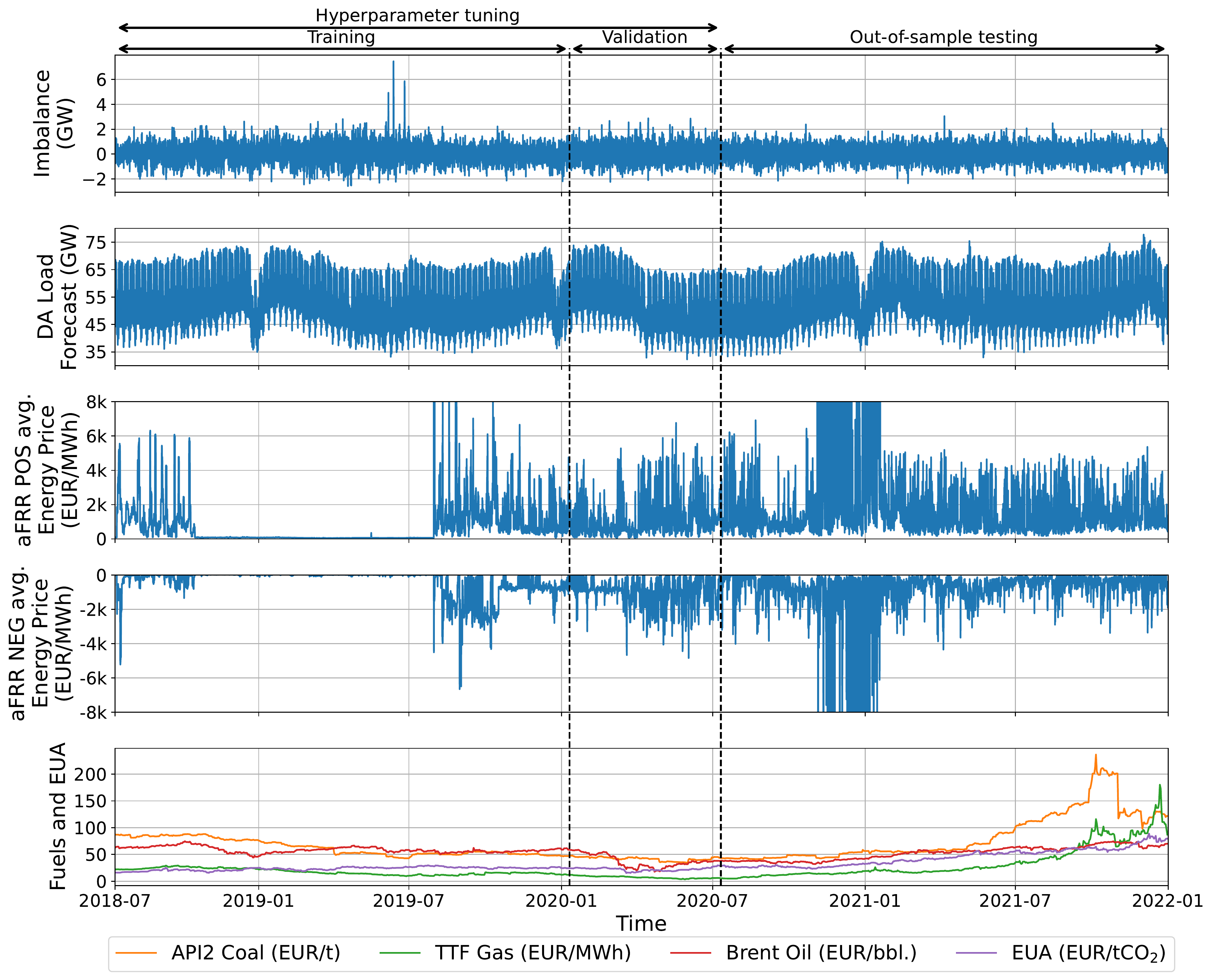}
	\caption{Time series plots of selected external regressors. POS and NEG stand for positive and negative, respectively.}
	\label{fig:data_tsplot}
	\end{figure}

	Figure~\ref{fig:data_tsplot} presents time series plots of selected external regressors and is a complement to Figure~\ref{fig:price_tsplot}.
	In both figures we marked the initial in-sample, the hyperparameter tuning, and the out-of-sample periods. Let us note the structural break in the aFRR positive and negative average energy prices between October 2018 and July 2019. During this period the mixed-pricing method was used in the tendering of aFRR and mFRR, i.e. both capacity and energy prices were used to select the cheapest BSPs. However, this method was abolished following a decision by the Düsseldorf Higher Regional Court due to an appeal and the capacity pricing (as described in Section~\ref{sec:balancing}) was immediately re-introduced.
	
	\section{Models and estimation}\label{sec:models}
	This section describes the input features and the models that use them to forecast the imbalance price $\text{IP}^{d,qh}$. In the EPF literature, it is typical to use autoregressive effects of the modelled prices, here however we cannot do it as the German imbalance prices are published once a month. For the price calculation in the IC market, we use the ${}_x\text{ID}_y$ definition of \citet{narajewski2020econometric}. Let us recall that the ${}_x\text{ID}_y$ is a volume-weighted average price of all transactions in the IC market that take place in the $[x+y, x)$ time interval prior the delivery.

	\subsection{Input features}\label{sec:input_features}
	The following features are considered in the exercise of modelling the imbalance price $\text{IP}^{d,qh}$ for day $d$ and quarter-hour $qh$ with $qh = 1, \dots, 96$. Whenever mentioning the corresponding product, we mean the same delivery hour, e.g. for $qh=6$ the corresponding hourly delivery time is 01:00 and quarter-hourly is 01:15. Note that we utilize only the information available until 30 minutes before the delivery.
	\begin{itemize}
		\itemsep0em
		\item Corresponding EPEX price indices: DA$^{d,h}$, IA$^{d,qh}$, ID$_1^{d,i}$, ID$_3^{d,i}$, and ID-Index$^{d,i}$ for $i = h,qh$ (8 regressors).\footnote{The ID-Index is a volume-weighted average price of all corresponding ID transactions.}
		\item Most recent 15-minute intraday prices ${}_x\text{ID}_{15\text{min}}^{d,h}$ for $h =1, \dots, 24$ and ${}_x\text{ID}_{15\text{min}}^{d,qh}$ for $qh = 1, \dots, 96$ ($24+96=120$ regressors).
		\item Corresponding intraday price differences $\Delta {}_{x}^{}\text{ID}_{5\text{min}}^{d,qh}$ with $x = 30, 35, \dots, 55$ min (6~regressors).
		\item DA forecasts of load, wind onshore, wind offshore and solar generation: $\text{Load}^{d,qh}$, $\text{WiOn}^{d,qh}$, $\text{WiOff}^{d,qh}$, $\text{Solar}^{d,qh}$ for $qh = 1, \dots, 96$ ($96 \cdot 4 = 384$ regressors).
		\item DA forecasts mentioned above for the previous day: $\text{Load}^{d-1,qh}$, $\text{WiOn}^{d-1,qh}$, $\text{WiOff}^{d-1,qh}$, $\text{Solar}^{d-1,qh}$ for $qh = 1, \dots, 96$ ($96 \cdot 4 = 384$ regressors).
		\item Most recent available imbalance values $\text{Imb}^{d,qh-i}$ for $i = 4, \dots, 7$ (4 regressors).
		\item aFRR prices: $\text{aFRR}_{i,j,k}^{d,qh}$ for $i = \text{POS}, \text{NEG}$ indicating the positive or negative balancing side, $j = \text{CAP}, \text{EN}$ indicating the capacity or energy price, and $k = \text{min}, \text{avg}, \text{max}$ indicating the minimum, average or maximum price ($6\cdot 2$ regressors).
		\item mFRR prices: $\text{mFRR}_{i,j,k}^{d,qh}$ with $i,j,k$ as above ($6\cdot 2$ regressors).
		\item Previous day coal, gas, oil and EUA prices: $\text{Coal}^{d-1}$, $\text{Gas}^{d-1}$, $\text{Oil}^{d-1}$, $\text{EUA}^{d-1}$ (4~regressors).
		\item Weekday dummies $\text{DoW}^d_i$ for $i = 1, \dots, 7$ (7~regressors).
		\item Cubic periodic B-splines $S^{d}_i$ for $i=1,\dots, 6$ constructed as in \citet{ziel2016forecasting2} (6~regressors).
	\end{itemize}

	In total, we consider 948 regressors for the modelling exercise. Let us shortly motivate the choice of these particular variables. Previous studies \cite{narajewski2020econometric, narajewski2020ensemble} have shown that the past prices can bring a lot of information regarding the future intraday price level and distribution. We expect similar behaviour in the imbalance price development, and thus we consider the price data, especially the most recent intraday prices and price differences. Similarly, the DA forecasts of fundamental variables might help in explaining the expected volatility. Naturally, the most recent intraday forecasts would be much more informative, but unfortunately this data is not publicly available and very expensive to obtain. The most recent observed imbalance values might indicate the expected imbalance in the considered quarter-hour. The aFRR and mFRR prices are natural regressors for the imbalance prices, as they directly contribute to their values. The fuel and EUA prices should explain the general price trend, and finally the weekday dummies and cubic B-splines account for weekly and annual seasonality, respectively.
	\subsection{Naive}
	Following the research on intraday markets \cite{narajewski2020econometric,narajewski2020ensemble, marcjasz2020beating,janke2019forecasting} where the authors find the most recent intraday price to be a very good and simple model, we construct the naive model in similar manner. That is to say, we assume the expected imbalance price to be equal the observed quarter-hourly ID$_1$ price
	\begin{equation}
		\mathbb{E}\left(\text{IP}^{d,qh}\right) = \text{ID}_1^{d,qh}.
	\end{equation}
	To obtain a distribution of imbalance prices, we use additionally the bootstrap method~\cite{efron1979bootstrap} which was successfully applied in previous EPF research studies \cite{narajewski2020ensemble, narajewski2021optimal, nowotarski2018recent, marcjasz2022distributional}. The in-sample bootstrapped errors are added to the forecasted expected price to derive the distribution forecast
	\begin{equation}
		\widehat{\text{IP}}^{D+1,qh}_{m} = \widehat{\mathbb{E}\left(\text{IP}^{D+1,qh}\right)} + \widehat{\varepsilon}_m^{D+1, qh} \, \, \text{for} \, \, m = 1, \dots, M
		\label{eq:bootstrap}
	\end{equation}
	where $\widehat{\varepsilon}_m^{D+1, qh}$ are drawn with replacement in-sample residuals for day $D+1$, i.e. we sample from the set of $\widehat{\varepsilon}^{d, qh} = \text{IP}^{d,qh} - \widehat{\text{IP}}^{d,qh}$ for $d = 1, \dots, D$.
	
	\subsection{Lasso with bootstrap}
	The lasso regression of \citet{tibshirani1996regression} is a very simple and powerful tool for linear model estimation, and thus gained high popularity and reputation in the EPF literature \cite{ziel2016forecasting, ziel2018day, lago2021forecasting, uniejewski2019understanding, narajewski2020econometric, marcjasz2020beating}. It serves both model estimation and variable selection, and therefore for the model we use all the regressors described in Section~\ref{sec:input_features}, and we denote such vector as $\bm{X}^{d,qh}$. The formula for the model is
	\begin{equation}
		\text{IP}^{d,qh} = \bm{X}^{d,qh}\bm{\beta}^{qh} + \varepsilon^{d,qh}
	\end{equation}
	and the lasso estimator is given by
	\begin{equation}
		\widehat{\bm{\beta}}^{qh} = \arg\min_{\bm{\beta}}\left\{\left|\left|\text{IP}^{d, qh} - \bm{X}^{d,qh}\bm{\beta}^{qh}\right|\right|_2^2 + \lambda ||\bm{\beta}||_1\right\}
		\label{eq:lasso_reg}
	\end{equation}
	where $\lambda$ is a tuning parameter. The lasso estimator expects scaled inputs, and in addition to that we apply on the inputs the variance stabilizing asinh transformation as suggested by \citet{uniejewski2017variance} with the inverse proposed by \citet{narajewski2020econometric}. The $\lambda$ parameter is tuned based on Bayesian information criterion (BIC) for $\lambda \in \Lambda = \{\lambda_i = 2^i | i \in \mathcal{G}\}$, where $\mathcal{G}$ is an equidistant grid from $-15$ to $1$ of length 50, similarly as in the paper of \citet{narajewski2020econometric}.
	Let us note that similarly as for the naive, the lasso model estimates the expected imbalance price and to obtain a distribution forecast we need the bootstrap procedure described in equation~\eqref{eq:bootstrap}.
	\subsection{Gamlss with lasso}
\begin{figure}[t!]
		\centering
		\includegraphics[width=\linewidth]{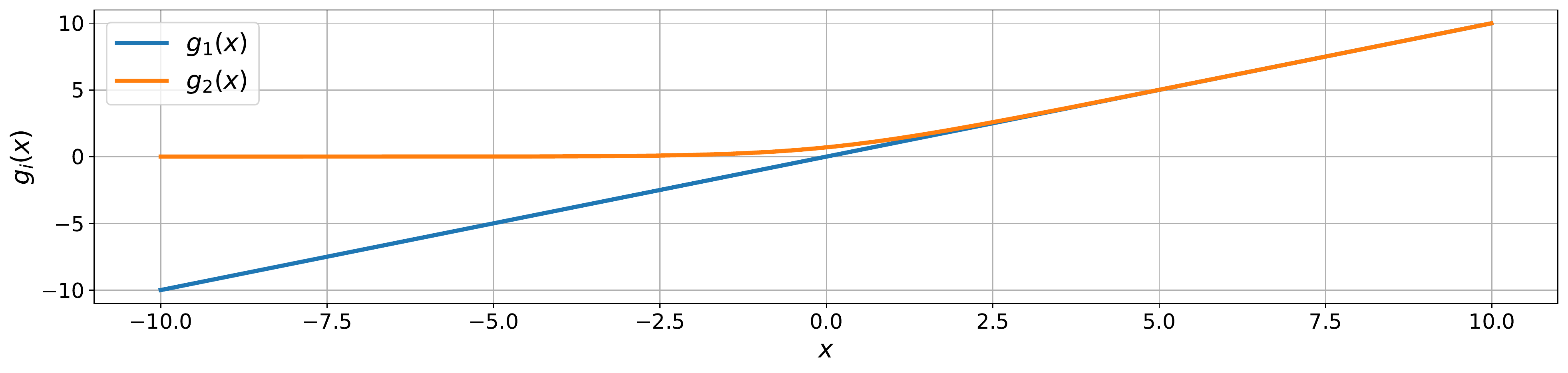}		
		\caption{Link functions used in the estimation of distribution parameters}
		\label{fig:links}
	\end{figure}

	The gamlss framework of \citet{rigby2005generalized} is an extension of the generalized additive models by allowing to build explicit additive models not only for the location, but also scale and shape parameters of a given distribution. Its potential was already noticed in the EPF literature \cite{gianfreda2018stochastic, narajewski2020ensemble}, however it has not gained yet such popularity as the lasso estimation. For the input vector $\bm{X}^{d,qh}$ we have the following model
	\begin{equation}
		g_i(\theta_i^{d,qh}) = \bm{X}^{d,qh}\bm{\beta}^{qh}_i
		\label{eq:gamlss}
	\end{equation}
	with $g_i$ being the link function, $\theta_i^{d,qh} \in \Theta^{d,qh}$ and distribution given by the cumulative distribution function $F(x;\Theta^{d,qh})$. 
	In the study, we consider the normal and t distributions. 
	The link function for the location parameter is the identity function $g_1(x) = x$, and for the scale and tail-weight the softplus function $g_2(x) = \log(\exp(x)+1)$. The link functions are shown in Figure~\ref{fig:links}.
	
	The model in equation~\eqref{eq:gamlss} is actually a glmlss one as we consider only linear effects of the inputs. Moreover, the size of $\bm{X}^{d,qh}$ could make the optimizing algorithm converge very slowly, especially for the 3-parametric distribution. Therefore, we additionally use the lasso regularization~\eqref{eq:lasso_reg} as described e.g. by \citet{ziel2021m5}, however we do not directly use the gamlss \cite{stasinopoulos2008generalized} and gamlss.lasso \cite{ziel2021extra} R packages as their deterministic algorithm has issues with convergence due to the very heavy tails of our data. Instead, we utilize the TensorFlow \cite{tensorflow2015-whitepaper} and Keras \cite{chollet2015keras} framework by building a simple neural network with a single linear hidden layer and given probability distribution as output. For each of the distribution parameters we use different regularization parameter $\lambda_i \in \left(10^{-5}, 10\right)$. We also allow for no regularization of each of the distribution parameters. The model is estimated by maximizing the log-likelihood using the Adam algorithm. The learning rate is assumed to be in the interval $(10^{-5}, 10^{-1})$, and we tune all the parameters using the Optuna \cite{akiba2019optuna} package in Python with the number of iterations arbitrarily set to 500. Depending on distribution, we have 5 or 7 hyperparameters to tune. Let us note that the input vector $\bm{X}^{d,qh}$ is standardized prior the modelling.

	\subsection{Probabilistic neural networks}
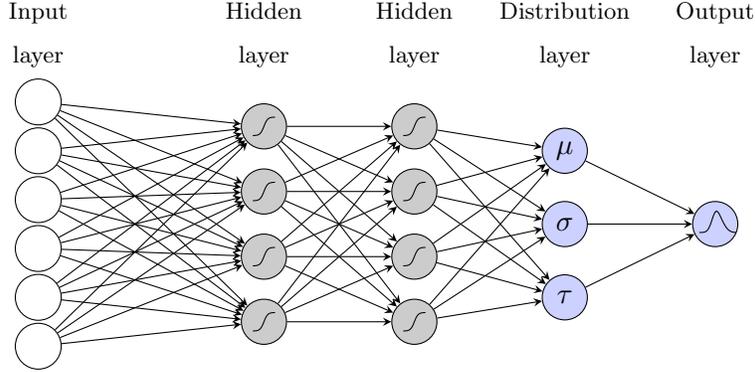
\begin{figure}[t!]
		\tikzset{%
			input neuron/.style={
				circle,
				draw,
				minimum size=.5cm
			},
			every neuron/.style={
				circle,
				draw,
				minimum size=.6cm
			},
			neuron missing/.style={
				draw=none, 
				scale=1.2,
				text height=.25cm,
				execute at begin node=\color{black}$\vdots$
			},
			sigmoid/.style={path picture= {
					\begin{scope}[x=.7pt,y=7pt]
						\draw plot[domain=-6:6] (\x,{1/(1 + exp(-\x))-0.5});
					\end{scope}
				}
			},
			linear/.style={path picture= {
					\begin{scope}[x=5pt,y=5pt]
						\draw plot[domain=-1:1] (\x,\x);
					\end{scope}
				}
			},
			mu/.style={path picture= {
					\begin{scope}[x=5pt,y=5pt]
						\draw node {${\mu}$};
					\end{scope}
				}
			},
			sigma/.style={path picture= {
					\begin{scope}[x=5pt,y=5pt]
						\draw node {${\sigma}$};
					\end{scope}
				}
			},
			tau/.style={path picture= {
					\begin{scope}[x=5pt,y=5pt]
						\draw node {${\tau}$};
					\end{scope}
				}
			},
			nu/.style={path picture= {
					\begin{scope}[x=5pt,y=5pt]
						\draw node {${\nu}$};
					\end{scope}
				}
			},
			gaussian/.style={path picture= {
					\begin{scope}[x=2pt,y=4pt]
						\draw plot[domain=-3:4] (\x,{1/sqrt(2*3.14)*(.5*exp(-\x*\x) +.5*exp(-(\x-.5)*(\x-1)/2))*5.5-.8});
					\end{scope}
				}
			},
		}
		\centering
		\begin{tikzpicture}[x=1cm, y=.65cm, >=stealth]
			
			\node [input neuron/.try, neuron 1/.try] (input-1) at (-1,2.5-1) {\makebox[8.5pt]{}};
			\node [input neuron/.try, neuron 2/.try] (input-2) at (-1,2.5-2) {\makebox[8.5pt]{}};
			\node [input neuron/.try, neuron 3/.try] (input-3) at (-1,2.5-3) {\makebox[8.5pt]{}};
			\node [input neuron/.try, neuron 4/.try] (input-4) at (-1,2.5-4) {\makebox[8.5pt]{}};
			\node [input neuron/.try, neuron 5/.try] (input-5) at (-1,2.5-5) {\makebox[8.5pt]{}};
			\node [input neuron/.try, neuron 6/.try] (input-6) at (-1,2.5-6) {\makebox[8.5pt]{}};
			
			\foreach \m [count=\y] in {1,2,3,4}
			\node [every neuron/.try, neuron \m/.try, fill= black!20, sigmoid ] (hidden-\m) at (2,7/3-\y*4/3) {};
			
			\foreach \m [count=\y] in {1,2,3,4}
			\node [every neuron/.try, neuron \m/.try, fill= black!20, sigmoid ] (hidden2-\m) at (4,7/3-\y*4/3) {};
			
			\foreach \m [count=\y] in {1}
			\node [every neuron/.try, neuron \m/.try, fill= blue!20, mu ] (par-1) at (6,2-1*1.5) {};
			
			\foreach \m [count=\y] in {1}
			\node [every neuron/.try, neuron \m/.try, fill= blue!20, sigma ] (par-2) at (6,2-2*1.5) {};
			
			\foreach \m [count=\y] in {1}
			\node [every neuron/.try, neuron \m/.try, fill= blue!20, tau ] (par-3) at (6,2-3*1.5) {};
			
			
			\foreach \m [count=\y] in {1}
			\node [every neuron/.try, neuron \m/.try, fill= blue!20, gaussian ] (output-\y) at (8,-1) {};
			
			
			\foreach \l [count=\i] in {1,2,3,4}
			\node [above] at (hidden-\i.north) {};
			
			
			\foreach \i in {1,...,6}
			\foreach \j in {1,...,4}
			\draw [->] (input-\i) -- (hidden-\j);
			
			\foreach \i in {1,...,4}
			\foreach \j in {1,...,4}
			\draw [->] (hidden-\i) -- (hidden2-\j);
			
			\foreach \i in {1,...,4}
			\draw [->] (hidden2-\i) -- (par-1);
			\foreach \i in {1,...,4}
			\draw [->] (hidden2-\i) -- (par-2);
			\foreach \i in {1,...,4}
			\draw [->] (hidden2-\i) -- (par-3);
			
			\foreach \i in {1,...,3}
			\foreach \j in {1}
			\draw [->] (par-\i) -- (output-\j);
			\foreach \l [count=\x from 1] in {Hidden, Hidden, Distribution, Output}
			\node [align=center, above] at (\x*2,2) {\footnotesize\l \\[-3pt] \footnotesize{layer}};
			
			\node [align=center, above] at (-1,2) {\footnotesize Input \\[-3pt] \footnotesize{layer}};
			
		\end{tikzpicture}	
		\caption{Exemplary network structure of the probabilistic MLP.}
		\label{fig:ann}
	\end{figure}

	The probabilistic neural network model is simply a multilayer perceptron (MLP) that models distribution parameters instead of price values, as shown in Figure~\ref{fig:ann}. Let us note that if we remove the hidden layers, we get the gamlss model described in the previous section. 
	 The approach of probabilistic MLP in EPF was first introduced by \citet{marcjasz2022distributional} for the day-ahead prices. For mathematical details see the aforementioned manuscript. The considered model assumes 2 or 3 hidden layers and outputs normal or t distribution. For the distribution parameters, we use the same link functions as in Figure~\ref{fig:links}. We regularize the model through input feature selection, $L_1$ regularization of the hidden layers and their weights, and a dropout layer.
	 We tune them together with the number of hidden layers, their activations functions, number of neurons and the learning rate. In the following, we present a list of all hyperparameters considered in the tuning.
	 \begin{itemize}
	 	\itemsep0em
	 	\item Input feature selection as described in Section~\ref{sec:input_features} (20~hyperparameters).
	 	\item Dropout layer -- whether to use the dropout layer after the input layer, and if yes at what rate. The rate parameter is drawn from $(0,1)$ interval (up to 2~hyperparameters).
	 	\item Size of the network -- either 2 or 3 hidden layers (1~hyperparameter).
 		\item Activation functions in the hidden layers. The possible functions are: elu, relu, sigmoid, softmax, softplus, and tanh (1~hyperparameter per layer).
 		\item Number of neurons in the hidden layers. The values are drawn from $[24,1024]$ interval (1~hyperparameter per layer).
	 	\item $L_1$ regularization -- whether to use the $L_1$ regularization on the hidden layers and their weights and if yes at what rate. The rate is drawn from $(10^{-5}, 10)$ interval (up to 4~hyperparameters per layer).
	 	\item Learning rate for the Adam algorithm drawn from $(10^{-5}, 10^{-1})$ interval (1~hyperparameter).
	 \end{itemize}
 
 	In total, we have up to 42 hyperparameters to tune. The selected input features are normalized prior the model estimation. Similarly as for the gamlss model, we use the Tensorflow \cite{tensorflow2015-whitepaper} and Keras \cite{chollet2015keras} framework for model estimation, and the Optuna \cite{akiba2019optuna} for hyperparameter tuning with the number of iterations arbitrarily set to 1000. The model contains additionally some elements which are not subject of the tuning exercise. These are size of the learning and validation sets, the optimizing algorithm, the number of epochs fixed to 1500, and batch size fixed to 32. We estimate the model by maximizing the log-likelihood, and we use the early stopping callback with patience of 50 epochs.
	
	\section{Application study}\label{sec:study}
	\subsection{Setting}
	Due to the high complexity of the models and the need for comprehensive and computational heavy hyperparameter tuning, we consider in the study only selected quarter-hours. That is to say, we use all quarter-hours of representative hours 0, 6, 12 and 18, i.e. $qh\in QH= \{1,2,3,4,25,26,27,28,49,50,51,52,73,74, 75, 76\}$. Such approach was already used in the literature. As described in Section~\ref{sec:models}, for each $qh$ we build separate models, including a separate hyperparameter tuning. Thus, we reduce the number of them from 96 to 16 without loss of generality. 
	
	The forecasting study utilizes a rolling window scheme with $D=730$ days in-sample and $N=539$ days out-of-sample. In case of gamlss and probabilistic MLP models, the in-sample period is split to $547$ days used for training and $183$ for validation. The hyperparameter tuning is performed once, using the initial in-sample data, as shown in Figures~\ref{fig:price_tsplot} and~\ref{fig:data_tsplot}. We aim for a very short-term forecasting utilizing the information available up to 30 minutes before the delivery. The naive and lasso models forecast the imbalance price distribution through $M=10000$ bootstrap samples, whereas the gamlss and probabilistic ANN models forecast directly the assumed distribution.
	
	\subsection{Evaluation}
	Following the conclusions of \citet{gneiting2007strictly}, our main evaluation measure is the continuous ranked probability score (CRPS) as it is a strictly proper scoring rule for marginal distribution forecasts. Additionally, we calculate the values of the RMSE, MAE and empirical coverage as supplementary measures. For statistically significant conclusions, we conduct the \citet{diebold1995comparing} test using the respective CRPS losses. In this subsection, we provide details regarding the calculation of the mentioned measures.
	
	The CRPS is approximated using the pinball loss
	\begin{equation}
		\text{CRPS}^{d,qh} = \frac{1}{R} \sum_{\tau \in r} \text{PB}^{d,qh}_{\tau}
	\end{equation}
	for a dense equidistant grid of probabilities $r$ between 0 and 1 of size $R$, see e.g.~\cite{nowotarski2018recent}. In our study, we consider $r = \{0.01, 0.02,\dots, 0.99\}$ of size $R = 99$.  $\text{PB}^{d,qh}_{\tau}$ is the pinball loss with respect to probability $\tau$. Its formula is given by
	\begin{equation}
		\text{PB}^{d,qh}_{\tau} = \left(\tau - \mathds{1}_{\left\{ {\text{IP}}^{d,qh} < \widehat{Q}_{\tau}^{d,qh}\right\}} \right) \left({\text{IP}}^{d,qh} - \widehat{Q}_{\tau}^{d,qh} \right)  
	\end{equation}
	where $\widehat{Q}_{\tau}^{d,qh}$ is a forecast of $\tau$-th quantile of $\text{IP}^{d,qh}$ price. To calculate the overall CRPS value we use a simple average
	\begin{equation}
		\text{CRPS} = \frac{1}{16N}   \sum_{qh \in QH} \sum_{d = 1}^{N} \text{CRPS}^{d,qh}.
	\end{equation}
	The formulas for the supplementary measures are given by
	\begin{equation}
		\tau\%\text{-cov} = \frac{1}{16N } \sum_{qh \in QH} \sum_{d = 1}^{N} \mathds{1}_{\left\{ \widehat{Q}^{d,qh}_{(1-\tau)/2} < \text{IP}^{d,qh} < \widehat{Q}^{d,qh}_{(1+\tau)/2} \right\} },
	\end{equation}
	\begin{equation}
		\text{RMSE}  =  \sqrt{\frac{1}{16N } \sum_{qh \in QH} \sum_{d = 1}^{N}\left({\text{IP}}^{d,qh} - \widehat{\mu}^{d,qh}\right)^2}
		\label{eq:rmse}
	\end{equation}
	and
	\begin{equation}
		\text{MAE} =  \frac{1}{16N } \sum_{qh \in QH} \sum_{d = 1}^{N} \left|{\text{IP}}^{d,qh} - \widehat{Q}_{0.5}^{d,qh} \right|
	\end{equation}
	where $\tau \in \{0.5, 0.9, 0.98\}$ and $\widehat{\mu}^{d,qh}$ is a forecast of expected $\text{IP}^{d,qh}$ price.
	
	The DM test measures the statistical significance of the difference between the accuracy of the forecasts of model $A$ and model $B$, and it is commonly used in the EPF literature \cite{narajewski2020econometric, narajewski2020ensemble, uniejewski2019understanding, ziel2018day}. Denote $L_Z^{d} = (L_Z^{d,qh})'_{qh\in QH}$ the vector of out-of-sample losses for day $d$ of model $Z$. Formally, we choose $L_Z^{d,qh} = \text{CRPS}^{d,qh}$.
	The multivariate loss differential series 
	\begin{equation}
		\Delta_{A,B}^d = ||L_A^d||_1 - ||L_B^d||_1
	\end{equation}
	defines the difference of losses in $||\cdot||_1$ norm. For each pair of models, we compute the p-value of two one-sided DM tests. The first one is with the null hypothesis $\mathcal{H}_0: \mathbb{E}(\Delta_{A,B}^d) \leq 0$, that is to say the outperformance of the forecasts of model $B$ by the ones of model $A$. The second test is with the reverse null hypothesis $\mathcal{H}_0: \mathbb{E}(\Delta_{A,B}^d) \geq 0$ and it complements the former one.
	
	\subsection{Results}
\begin{table}[b!]
\centering
\begin{tabular}{rrrrrrr}
  \hline
 & CRPS & MAE & RMSE & 50\%-cov & 90\%-cov & 98\%-cov \\ 
  \hline
Naive & \cellcolor[rgb]{0.531,0.91,0.5} {23.04} & \cellcolor[rgb]{0.5,0.9,0.5} {\textbf{61.22}} & \cellcolor[rgb]{0.5,0.9,0.5} {\textbf{115.2}} & \cellcolor[rgb]{1,0.5,0.55} {0.2847} & \cellcolor[rgb]{1,0.5,0.542} {0.8011} & \cellcolor[rgb]{1,0.75,0.5} {0.9525} \\ 
  Lasso & \cellcolor[rgb]{1,1,0.5} {24.91} & \cellcolor[rgb]{0.963,1,0.5} {65.06} & \cellcolor[rgb]{1,0.948,0.5} {124.5} & \cellcolor[rgb]{1,0.5,0.55} {0.2784} & \cellcolor[rgb]{1,0.5,0.55} {0.7825} & \cellcolor[rgb]{1,0.606,0.5} {0.9417} \\ 
  gamlss.N & \cellcolor[rgb]{1,0.5,0.55} {36.09} & \cellcolor[rgb]{1,0.5,0.55} {88.70} & \cellcolor[rgb]{1,0.5,0.55} {154.5} & \cellcolor[rgb]{1,0.546,0.5} {0.3294} & \cellcolor[rgb]{1,0.5,0.55} {0.6158} & \cellcolor[rgb]{1,0.5,0.55} {0.7067} \\ 
  gamlss.t & \cellcolor[rgb]{0.866,1,0.5} {24.24} & \cellcolor[rgb]{1,0.994,0.5} {65.70} & \cellcolor[rgb]{1,0.952,0.5} {124.2} & \cellcolor[rgb]{1,0.799,0.5} {0.4074} & \cellcolor[rgb]{0.846,1,0.5} {\textbf{0.8851}} & \cellcolor[rgb]{0.854,1,0.5} {0.9739} \\ 
  probNN.N & \cellcolor[rgb]{1,0.5,0.55} {35.67} & \cellcolor[rgb]{1,0.5,0.55} {92.84} & \cellcolor[rgb]{1,0.5,0.55} {368.5} & \cellcolor[rgb]{0.572,0.924,0.5} {\textbf{0.5037}} & \cellcolor[rgb]{1,0.768,0.5} {0.8460} & \cellcolor[rgb]{1,0.5,0.55} {0.9249} \\ 
  probNN.t & \cellcolor[rgb]{1,0.951,0.5} {25.85} & \cellcolor[rgb]{1,0.922,0.5} {68.74} & \cellcolor[rgb]{1,0.869,0.5} {129.2} & \cellcolor[rgb]{1,0.747,0.5} {0.3912} & \cellcolor[rgb]{1,0.957,0.5} {0.8733} & \cellcolor[rgb]{1,0.911,0.5} {0.9645} \\ 
  Combination & \cellcolor[rgb]{0.5,0.9,0.5} {\textbf{22.94}} & \cellcolor[rgb]{0.731,0.977,0.5} {62.84} & \cellcolor[rgb]{0.755,0.985,0.5} {117.8} & \cellcolor[rgb]{1,0.606,0.5} {0.3480} & \cellcolor[rgb]{1,0.968,0.5} {0.8749} & \cellcolor[rgb]{0.5,0.9,0.5} {\textbf{0.9787}} \\ 
   \hline
\end{tabular}
\caption{Error measures of the considered models. Colour indicates the performance columnwise (the greener, the better). With bold, we depicted the best values
	in each column.}
\label{tab:results}
\end{table}

	Table~\ref{tab:results} presents the results of the forecasting study. We see that the lowest error values are obtained for the naive model which forecasts the imbalance price simply with the quarterly ID$_1$ price. However, its empirical coverage is very bad.
	The second-lowest errors are produced by the gamlss model that assumes the t-distribution, and this model provides the best values in terms of 90\% and 98\% empirical coverage. The generalization from gamlss to probabilistic neural network model does not bring any improvement for the t-distribution. 
	Based on the performance of the two mentioned models, we decided to try a simple forecast combination by averaging the forecasts of the two models. This brings a small improvement in the CRPS and in the coverage, compared to the naive model. 
	Let us also mention very high errors of the models that assume the normal distribution. This is inline with the previous studies \cite{narajewski2020ensemble} on intraday price development, and it was expected based on Figures~\ref{fig:price_tsplot} and~\ref{fig:price_histogram}. Interestingly, the probNN.N model provides a very accurate 50\% coverage, but not as good 90\% or 98\% coverages. Finally, the lasso model performs slightly worse than the naive in all terms what indicates that one cannot gain any improvement only with linear terms.
	
	\begin{figure}[b!]
	\centering
	\includegraphics[width=\linewidth]{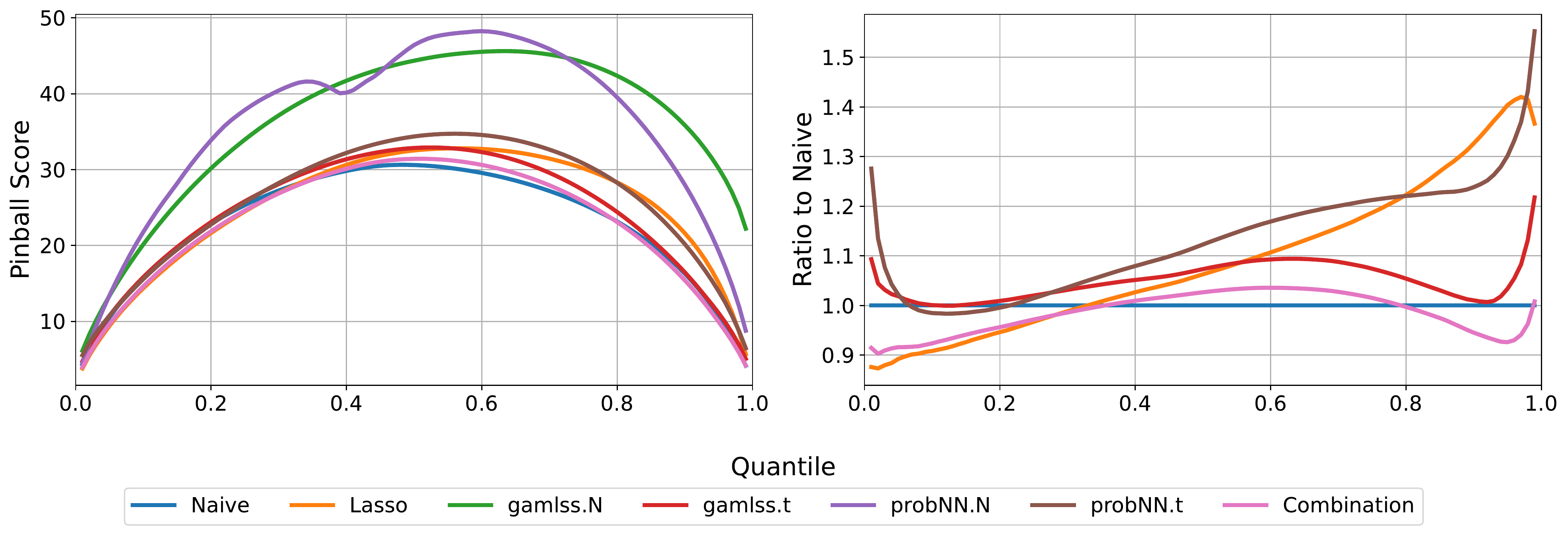}
	\caption{Pinball score (left) and its ratio to the naive (right) over quantiles $\tau \in r$. The right graph shows selected models for better clarity.}
	\label{fig:PS_over_quantiles}
	\end{figure}
	\begin{figure}[b!]
	\centering
	\includegraphics[width=\linewidth]{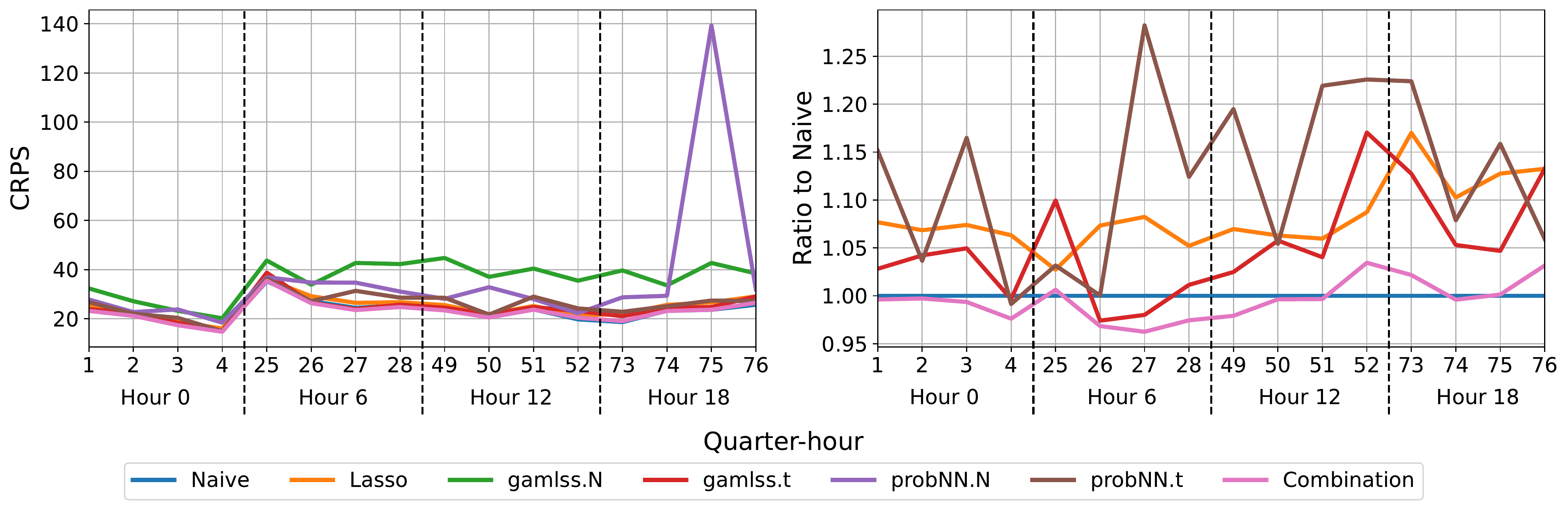}
	\caption{CRPS (left) and its ratio to the naive (right) over quarter-hours $qh \in QH$. The right graph shows selected models for better clarity.}
	\label{fig:PS_over_hours}
	\end{figure}

	Figure~\ref{fig:PS_over_quantiles} shows the pinball score values over quantiles $\tau \in r$ and the ratio to the naive model. For better clarity, we removed from the right plot the models assuming normal distribution. We see that the models have generally more issues with forecasting the right tail of the distribution. Interestingly, the lasso model forecasts the quantiles up to around 0.3 slightly better than the naive, however it loses very much in the higher quantiles. Also, the combination of naive and gamlss.t is slightly better than the naive in both tails, however not that good in the central part of the distribution. This shows that a forecast combination, as e.g. in \citet{berrisch2021crps}, could likely improve the overall score. Figure~\ref{fig:PS_over_hours} presents the CRPS values over considered quarter-hours. The naive model seems to be the best across all quarter-hours except for two at hour 6. There, the gamlss.t is slightly better than the naive, however the difference is not large and probably not significant. Again, some additional improvement comes as a result of combining the naive with the gamlss.t model.
	
	\begin{figure}[t!]
		\centering
		\input{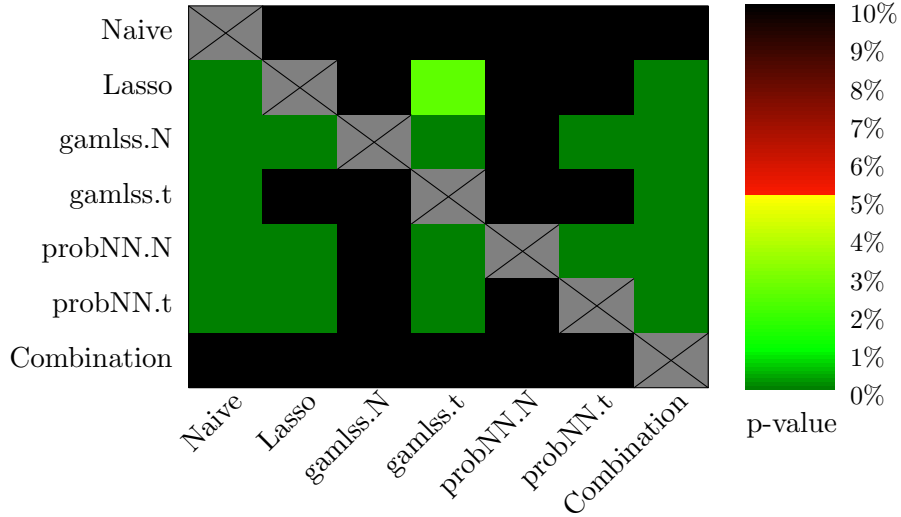}
		\caption{Results of the Diebold-Mariano test. The plots present p-values for the CRPS$^{d,qh}$ loss --- the closer they are to zero ($\to$ dark green), the more significant the difference is between forecasts of X-axis model (better) and forecasts of the Y-axis model (worse).}
		\label{fig:dm_test}
	\end{figure}

	Finally, Figure~\ref{fig:dm_test} provides p-values of the DM test obtained using CRPS loss. This figure only confirms the conclusions that we made based on Table~\ref{tab:results}. Namely, the forecasts of the naive model are significantly the best among considered models and the ones of gamlss.t model the second-best. Moreover, the combination of naive and gamlss.t is not significantly different to the naive itself.
 
	\section{Conclusion}\label{sec:conclusion}
	The paper raised the novel issue of probabilistic imbalance electricity price forecasting in the German market. The participation in the balancing is mandatory for every market player, and therefore this subject is crucial for them. The analysis assumed a setting of a very short-term forecasting, 30 minutes before the physical delivery. We considered various state-of-art methods for probabilistic EPF, however none of them could provide better forecasts in terms of CRPS, MAE and RMSE than the naive ID$_1$ price. On the other hand, the gamlss and probabilistic neural networks models provide forecasts with far higher empirical coverage than the naive. This is an evidence that the results might be improved, e.g. using intraday power generation forecasts or forecasting combination methods, e.g. \cite{berrisch2021crps}.
	
	The obtained results are an argument towards the market efficiency between the intraday and balancing markets. This extends the conclusions of intraday market being close to market efficiency \cite{narajewski2020econometric, marcjasz2020beating}. Therefore, given the difficulty in forecasting the imbalance prices and the potential size of forecasting errors, the BRPs should minimize their imbalance rather than seeking opportunities in the balancing market.

	\section*{Acknowledgments}
	This research article was partially supported by the German Research Foundation (DFG, Germany) and the National Science Center (NCN, Poland) through BEETHOVEN grant no. 2016/23/G/HS4/01005.
	
\vspace{-5mm} 
\bibliographystyle{unsrtnat1}

\bibliography{bibliography}

\end{document}